\documentclass[a4paper]{jpconf}
\usepackage{amsmath}
\usepackage{graphicx}

\newcommand{\sNN}{{{$\sqrt{s_{_{{NN}}}}$}}}

\newcommand{\muB}{\mbox{$\mu_{B}$}}

\newcommand{\KV}{{\mbox{$\kappa\sigma^{2}$}}}
\newcommand{\SD}{{\mbox{$S\sigma$}}}

\begin{document}
\title{Volume Fluctuation and Autocorrelation Effects in the Moment Analysis of Net-proton Multiplicity Distributions in
Heavy-Ion Collisions}

\author{Xiaofeng Luo}\address{Key Laboratory of Quark and Lepton Physics (MOE) and Institute of Particle Physics, Central China Normal University, Wuhan 430079, China}
\author{Ji Xu}\address{Key Laboratory of Quark and Lepton Physics (MOE) and Institute of Particle Physics, Central China Normal University, Wuhan 430079, China}
\author{Bedangadas Mohanty}\address{National Institute of Science Education and Research, Bhubaneswar 751005, India}
\author{Nu Xu}\address{Key Laboratory of Quark and Lepton Physics (MOE) and Institute of Particle Physics, Central China Normal University, Wuhan 430079, China}
\address{Nuclear Science Division, Lawrence Berkeley National Laboratory, Berkeley, CA 94720, USA.}

\date{\today}
\ead{xfluo@iopp.ccnu.edu.cn}

\begin{abstract}
Moments (Variance ($\sigma^2$), Skewness($S$), Kurtosis($\kappa$)) of multiplicity distributions of conserved quantities, such as net-baryon,net-charge and net-strangeness, are predicted to be sensitive to the correlation length of the system and connected to the thermodynamic susceptibilities computed in Lattice QCD and Hadron Resonance Gas (HRG) model. In this paper, we present several measurement artifacts that could lead to volume fluctuation and auto-correlation effects in the moment analysis of net-proton multiplicity distributions in heavy-ion collisions using the UrQMD model. We discuss methods to overcome these artifacts so that the extracted moments could be used to obtain physical conclusions. In addition we present methods to properly estimate the statistical errors in moment analysis.
\end{abstract}

\section{Introduction}
\label{sect_intro}

One of the main goals of performing heavy-ion collision experiments is to explore the phase structure of the QCD matter~\cite{bes}. The first principle Lattice QCD calculation shows that the transition from hadronic to partonic matter at zero {\muB} is a smooth crossover~\cite{crossover}, while at finite {\muB} region there could be a first order phase transition~\cite{firstorder}. The end point of the first order phase transition boundary is the so called QCD Critical Point (CP)~\cite{qcp}. It is well known that there are large uncertainties in the current Lattice QCD calculation for determining the first order phase boundary as well as the QCD critical point~\cite{location} in the QCD phase diagram. This is due to the sign problem at finite {\muB} region~\cite{methods}.

In heavy-ion collision, the moments (Variance ($\sigma^2$), Skewness($S$), Kurtosis($\kappa$)) of distributions of conserved quantities, such as net-baryon, net-charge and net-strangeness, are predicted to be sensitive to the correlation length of hot dense matter created in the collisions~\cite{qcp_signal} and connected to the thermodynamic susceptibilities computed in Lattice QCD~\cite{qcp,MCheng2009,Wupp_Lattice} and in the Hadron Resonance Gas (HRG)~\cite{HRG} model. For instance, the higher order cumulants of multiplicity ($N$) distributions are proportional to the high power of the correlation length ($\xi$) as third order cumulant $C_{3}=S \sigma^{3}=<(\delta N)^3> \sim \xi^{4.5}$ and fourth order cumulant $C_{4}=\kappa \sigma^{4}=<(\delta N)^4> - 3(\delta N)^2> \sim \xi^{7}$~\cite{qcp_signal}, where the $\delta N=N-<N>$ and $<N>$ is the mean multiplicities. The moment products, {\KV} and {\SD}, are also related to the ratios of various order susceptibilities, such as ratios of baryon number susceptibilities can be compared with the experimental data as $\kappa \sigma^2=\chi^{(4)}_{B}$/$\chi^{(2)}_{B}$ and $S
\sigma=\chi^{(3)}_{B}$/$\chi^{(2)}_{B}$~\cite{science}. Those susceptibility ratios cancel the volume effect to first order. Due to the sensitivity to the correlation length and the connection with the susceptibilities, we can use the higher moments of the conserved quantity distributions (within certain phase space) to search for the QCD critical point and map the first order phase boundary in the QCD phase diagram~\cite{QM2012_Xiaofeng}. 

As the moment analysis of conserved quantities is a powerful observable to explore the QCD phase structure and study the phase transition properties of QCD matter, these have been widely studied experimentally and theoretically. To precisely extract physics message from experimental data, it is very important to address the background and artifact effects, such as volume fluctuation and auto-correlation effects, in the moment measurement. In this paper, we have developed several techniques to address those effects in the moment of net-proton multiplicity distributions, which will be demonstrated by using the events from a transport model. These techniques are not restricted to the net-proton analysis and can be also applied to the moments analysis of the net-charge and net-strangness, which are also being studied at RHIC. It will also provide a reference on how to perform model calculation for comparison with the experimental results.  

The paper is organized as follows. In section II we briefly discuss the UrQMD model used for the analysis presented here. In section III we define the observables. In section IV, we present different methods of estimating the statistical errors for moment analysis. Sections V and VI discuss the measurement artifacts, centrality bin width effect (CBWE), centrality resolution effect (CRE) and Auto-correlations effect (ACE), respectively. 
Finally in section VII we present a summary of the work.

\section{UrQMD Model}
\label{UrQMD}

The Ultra Relativistic Quantum Molecular Dynamics (UrQMD) is a microscopic many-body approach to study $p + p$, $p + A$, and $A + A$ interactions at relativistic energies and is based on the covariant propagation of color strings, constituent quarks, and diquarks accompanied by mesonic and baryonic degrees of freedom. Furthermore it includes rescattering of particles, the excitation and fragmentation of color strings, and the formation and decay of hadronic resonances. UrQMD is a transport model for simulating heavy-ion collisions in the energy range from SIS to RHIC (even in LHC). It combines different reaction mechanism, and can provide theoretical simulated results of various experimental observables. The main parts of UrQMD model are: two body reaction cross section, two body potential and  decay width. More details about the UrQMD model can be found in the reference~\cite{urqmd}.

We performed our calculations with UrQMD model for
Au+Au collisions at \sNN = 7.7, 11.5, 19.6, 27, 39, 62.4, 200 GeV and the corresponding event statistics are 107, 108, 50, 65, 93, 29 and 28 million, respectively. We measure event-by-event wise $N_{p-\bar{p}}$ at the mid-rapidity ($|y|<0.5$) and within the transverse momentum $0.4<p_{T}<0.8$ GeV/c. This is same kinematic range used in moment analysis of net-proton distributions as in the STAR experiment~\cite{PRL,QM2012_Xiaofeng,WWND2011,SQM2011}.

\section{Observables}
\label{sect_obs}

In statistics, probability density distribution functions can be
characterized by the various moments, such as mean ($M$), variance
($\sigma^2$), skewness ($S$) and kurtosis ($\kappa$). Before
introducing the above moments used in our analysis, we would like to
define cumulants, which are alternative approach compared to moments to characterize a distribution. 
The cumulants determine the moments in the sense that
any two probability distributions whose cumulants are identical will
have identical moments as well, and similarly the moments can determine
the cumulants.

In the following, we use
$N$ to represent the net-proton number $N_{p-\bar{p}}$ in one event.
The average value over whole event ensemble is denoted by $<N>$,
where the single angle brackets are used to indicate ensemble
average of a event-by-event distribution.

The deviation of $N$ from its mean value is defined by

\begin{equation}
  \delta N=N-<N>
\end{equation}

The various order cumulants of event-by-event
distributions of a variable $N$ are defined as:

\begin{eqnarray}
C_{1,N}&=&<N> \\
C_{2,N}&=&<(\delta N)^{2}> \\
C_{3,N}&=&<(\delta N)^{3}> \\
C_{4,N}&=&<(\delta N)^{4}>-3<(\delta N)^{2}>^{2}
\end{eqnarray}

An important property of the cumulants is their additivity for
independent variables. If $X$ and $Y$ are two independent random
variables, then we have $C_{i,X+Y}=C_{i,X}+C_{i,Y}$ for $i$th order
cumulant. 

Once we have the definition of cumulants, various moments can be
obtained as:
\begin{eqnarray}
M=C_{1,N},\sigma^{2}=C_{2,N},S=\frac{C_{3,N}}{(C_{2,N})^{3/2}},\kappa=\frac{C_{4,N}}{(C_{2,N})^{2}}
\end{eqnarray}

And also, the moments product $\kappa \sigma^{2}$ and $ S \sigma$
can be expressed in terms of the ratio of cumulants like: 

\begin{eqnarray}
\kappa \sigma^{2}=\frac{C_{4,N}}{C_{2,N}},
S\sigma=\frac{C_{3,N}}{C_{2,N}}
\end{eqnarray}

With above definition, we can calculate various
moments and moment products from the measured event-by-event
net-proton distribution for each centrality.

\section{Statistical Error Estimation}
\label{error estimation}

As the moment analysis is a statistics hungry study, the error estimation is crucial to extract physics information from the limited experimental data. We will present several statistical methods (Delta theorem, Bootstrap~\cite{bootstrap}, Sub-group) of error estimation in the moment analysis and compare them through a Monte Carlo simulation. For simplicity, we
assume the particles are independently distributed as Poissonian distributions. Then, the difference of two independent Poisson
distributions is distributed as "Skellam" distribution. The skellam probability density distribution is

\begin{eqnarray}
f(k;\mu_1,\mu_2)=e^{\mu_1 + \mu_2}(\frac{\mu_1}{\mu_2})^{k/2}I_{|k|}(2\sqrt{\mu_1 \mu_2})
\end{eqnarray}

, where the $\mu_1$ and $\mu_2$ are the mean value of the Poisson distributions of particle type 1 and 2, respectively, the $I_k (z)$
is the modified bessel function of the first kind. Then, we can calculate various moments ($M,\sigma,S,\kappa$) and moment products ($\kappa\sigma^2, S\sigma$) products of the Skellam distribution. The relations 
are given below:

\begin{eqnarray}
M=\mu_1 - \mu_2,
\sigma=\sqrt{\mu_1 + \mu_2} \\
S=\frac{\mu_1 - \mu_2}{(\mu_1 + \mu_2)^{2/3}} ,
\kappa=\frac{1}{\mu_1 + \mu_2} \\
S\sigma=\frac{\mu_1 - \mu_2}{\mu_1 + \mu_2} , \kappa\sigma^2=1
\end{eqnarray}

To do the simulation, we set the two mean values of the Skellam distributions as $\mu_1$ = 4.11, $\mu_2$ = 2.99,
which are similar with the mean proton and anti-proton numbers measured in central
Au+Au collisions at \sNN = 200 GeV by the STAR experiment~\cite{PRL}. Then, we generate
random numbers as per the Skellam distribution. The details of Delta theorem error estimation method for moment analysis can be found in~\cite{Delta_theory}.
The bootstrap method~\cite{bootstrap} is based on the repeated sampling with the same statistics of the parent distribution and the statistical errors can be obtained as the root mean square of distribution of the observable from each sample. In the sub-group method, we randomly divide the whole sample into several sub-groups with same statistics and the errors 
are obtained as the root mean square of the distribution of the observable from each sub-group. In the simulation presented in the paper, we set 200 bootstrap times and 5 sub-groups.

\begin{figure}[thpb]
\centering
\renewcommand{\figurename}{FIG.}
\vskip 0cm

\includegraphics[width=0.7\textwidth]{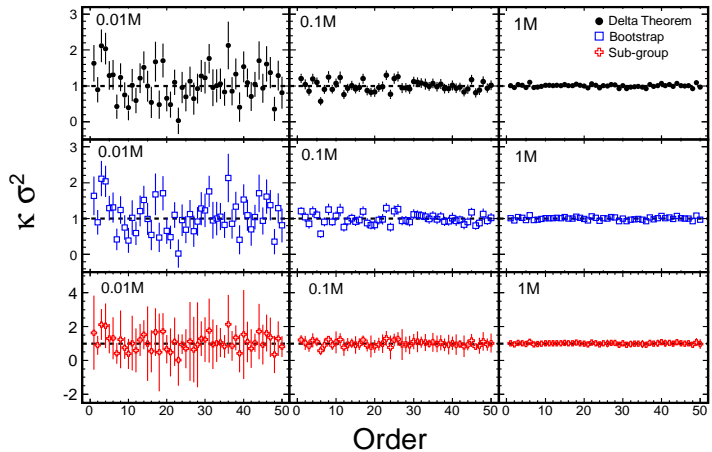}
\caption{(Color online) $\kappa \sigma^{2}$ of 50 samples arranged by order, which are independently and randomly generated from Skellam distribution with different number of events (0.01, 0.1, 1 million). The dashed lines are expectations value 1.}
\label{Plot::error}
\end{figure}

Figure~\ref{Plot::error} shows the error estimation comparison between Delta theorem, Bootstrap and Sub-group methods for observable value ($\kappa \sigma^{2}$) of the Skellam distribution constructed in this study. For each method, fifty independent sets are sampled from the Skellam distribution with 
statistics 0.01 , 0.1 and 1 million, respectively. The probability for the observable 
value  ($\kappa \sigma^{2}$) staying within $\pm 1\sigma$ around expectation is about $68.3 \%$. That would lead to error bars of about 33 out of 50 points should touch the expected value(dashed line at unity) in Fig.~\ref{Plot::error}. We find that the results from the Delta theorem and Bootstrap method show similar error values and satisfies the above statistical criteria. It indicates that the Delta theorem and Bootstrap error estimation methods for the moment analysis are reasonable and can reflect the statistical properties of moments. while there are 49 out of 50 error bars touch the expected value in sub-group method, which means the sub-group method overestimates the statistical errors.

According to the error formula derived from delta theorem, for normal distributions, the errors for
cumulants ratios are proportional to the standard deviation of the distribution as:
\begin{eqnarray}
error(\hat{S} \hat{\sigma}) \propto \frac{\sigma}{\sqrt{n}} \\
error(\hat{\kappa}\hat{\sigma}^2) \propto \frac{\sigma^2}{\sqrt{n}}
\end{eqnarray}, where the $n$ is the number of events.

\section{Volume Fluctuation Effect}
In heavy-ion collisions, the collision centrality and/or initial collision geometry of two nuclei with finite volume is estimated by comparing the particle multiplicities with Glauber simulation~\cite{Glauber}
and can not be measured directly. This drawback in general can cause two effects in the moment analysis of particle multiplicity
distributions within finite centrality bin. One is the so called centrality bin width effect~\cite{WWND2011}, which is caused by volume variation within a finite centrality bin size and the other one is centrality resolution effect, which is caused by the initial volume fluctuations~\cite{VolumeFluctuation}. In the following two sub-sections, we will discuss those two effects caused by volume fluctuation in detail with UrQMD model simulation and the
corresponding methods to address those effects.

\subsection{Centrality Bin Width Effect (CBWE) and Correction}
\label{cbwc}
In heavy ion-collisions, the collision centrality can be explained as a percentage of the total cross-section, such as 0-5\% (most central) and 30-40\% (semi-peripheral), which indicates the fraction of a data sample relative to all possible collision geometries. 
It is usually determined by an observable like the charged particle multiplicity, in which the smallest centrality bin is a single multiplicity value. Generally, we report the results for a wider centrality bin, such as 0-5\% and 5-10\%, to have better 
statistical accuracy. But, before calculating various moments of particle number distributions for one
wide centrality bin, such as 0-5\%, 5-10\%, we should consider the so called Centrality Bin Width Effect (CBWE) arising from the impact
parameter (or volume) variations due to the finite centrality bin. This effect
must be eliminated, as an artificial centrality dependence could be introduced due to finite centrality bin width.
To demonstrate this effect, we define the centrality by the charged kaon and pion 
multiplicities within $|\eta|<2$ in the UrQMD model calculations. This centrality definition is to avoid the auto-correlation and centrality resolution effects in the net-proton moment analysis, which will be discussed later.
Before calculating various moments of particle number distributions for one
wide centrality bin, such as 0-5\%, 5-10\%, we should consider the so
called Centrality Bin Width Effect (CBWE) arising from the impact
parameter (or volume) variations due to the finite centrality bin. This effect
must be eliminated, as an artificial centrality dependence could be introduced due to finite centrality bin width.
\begin{figure}[hptb]
\centering
\renewcommand{\figurename}{FIG.}
\vskip 0cm
\includegraphics[width=0.7\textwidth]{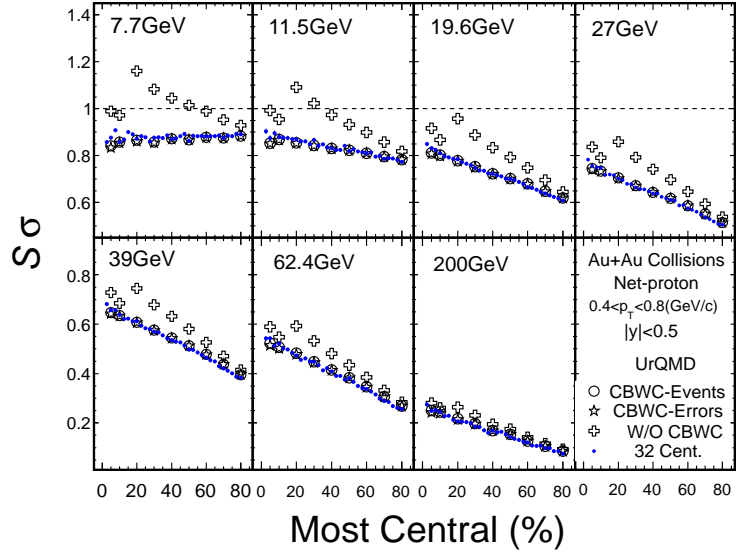}
\caption{(Color online) The centrality dependence of the moments
products $(S\sigma)$ of net-proton multiplicity distributions for Au+Au collisions at \sNN=7.7, 11.5, 19.6, 27, 39, 62.4, 200GeV in UrQMD model. The solid dots represent the results calculated from 32 centrality bins.}
\label{Plot::CBWC-sd}
\end{figure}

\begin{figure}[hptb]
\centering
\renewcommand{\figurename}{FIG.}
\vskip 0cm
\includegraphics[width=0.7\textwidth]{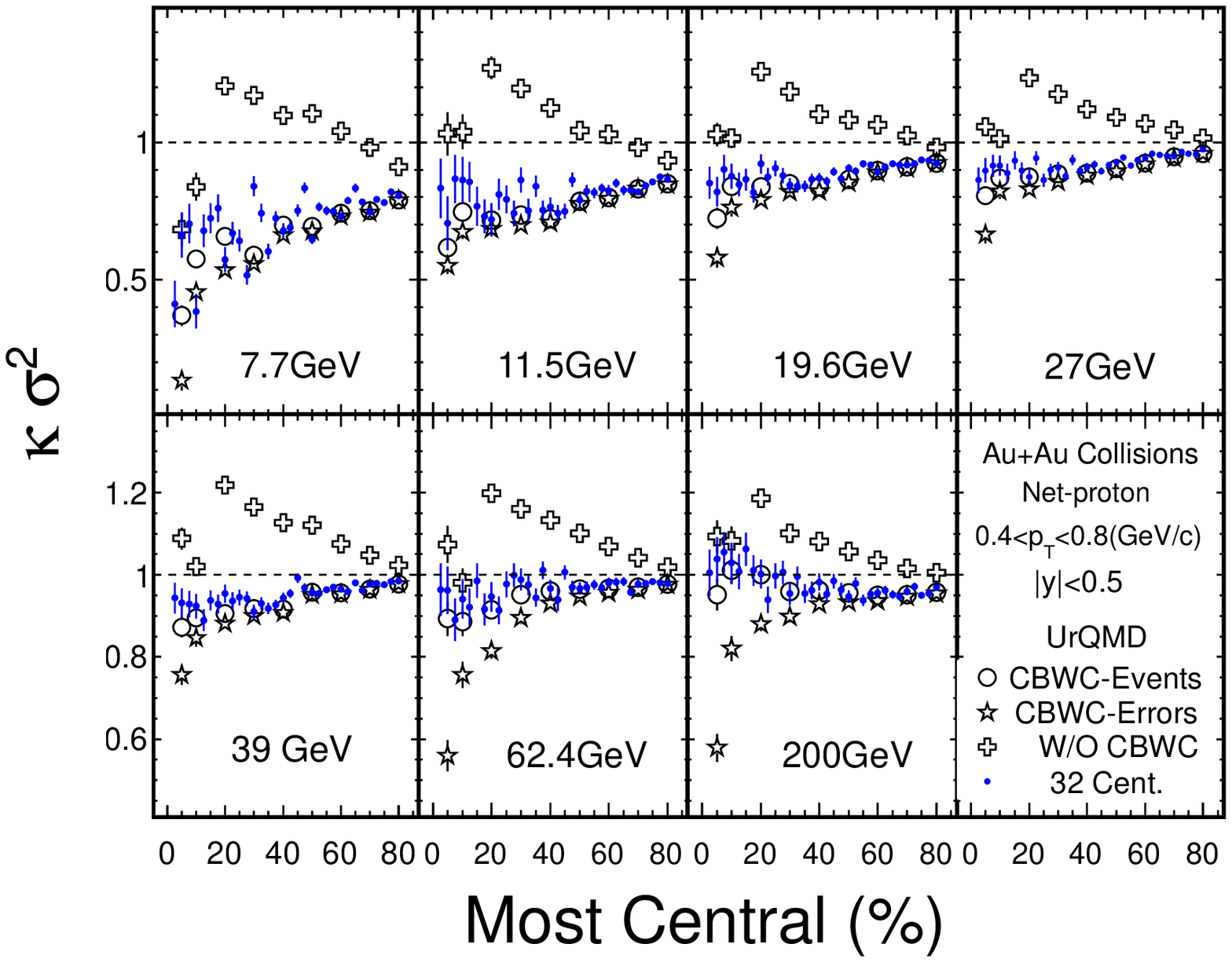}
\caption{(Color online) The centrality dependence of the moments
products $(\kappa\sigma^2)$ of net-proton multiplicity distributions for Au+Au collisions at \sNN=7.7, 11.5, 19.6, 27, 39, 62.4, 200GeV in UrQMD model. The solid dots represent the results calculated from 32 centrality bins}
\label{Plot::CBWC-kv}
\end{figure}

To eliminate the centrality bin width effect, we have developed a technique called 
Centrality Bin Width Correction (CBWC), to calculate the various moments of particle number distributions 
in one wide centrality bin. Those formulas are:

\begin{eqnarray}
\sigma  &=& \frac{{\sum\limits_r^{} {n_r \sigma _r  }
}}{{\sum\limits_r^{} {n_r } }} = \sum\limits_r^{} {\omega _r \sigma
_r  }    \\
S &=& \frac{{\sum\limits_r^{} {n_r S_r } }}{{\sum\limits_r^{} {n_r }
}} = \sum\limits_r^{} {\omega _r } S_r \\
\kappa  &= &\frac{{\sum\limits_r^{} {n_r \kappa _r }
}}{{\sum\limits_r^{} {n_r } }} = \sum\limits_r^{} {\omega _r }
\kappa _r
\end{eqnarray}

, where the $n_r$ is the number of events in $r^{th}$ multiplicity for centrality determination, the $\sigma_{r}$, $S_{r}$ and $\kappa_{r}$ represent the
standard deviation, skewness and kurtosis of particle number distributions at $r^{th}$ multiplicity. The
corresponding weight for the $r^{th}$ multiplicity is $\omega_r={{ {n_r  } }}/{{\sum\limits_r^{} {n_r
} }}$.

Figure~\ref{Plot::CBWC-sd} and~\ref{Plot::CBWC-kv} show the centrality dependence of the moment
products $(S\sigma , \kappa\sigma^2)$ of net-proton multiplicity distributions for Au+Au collisions at \sNN=7.7, 11.5, 19.6, 27, 39, 62.4 and 200 GeV obtained from the UrQMD model simulation. The open circle and open cross in Fig.~\ref{Plot::CBWC-sd} and~\ref{Plot::CBWC-kv} represent the results calculated with 
and without the technique of CBWC in the nine centralities (0-5\%, 5-10\%, 10-20\%, 20-30\%...70-80\%), respectively. The
solid circles show the result of centrality dependence of 32 bins in centrality (0-2.5\%, 2.5-5\%, 5-7.5\%...77.5\%-80\%), in which the 
CBWC is not applied. For the case of nine centralities, we clearly observe that the results
with CBWC are very different from those without CBWC. It means the CBWE do have a significant contribution to the value of the moment analysis and the CBWC will make the value of the moment products smaller
by reducing the effect of variation of volume in one wide centrality bin. On the other hand, the results calculated from 32 centrality bins, for which the CBWE is small, are good agreement with the results from nine centralities with CBWC, which confirms the effectiveness of the proposed CBWC method. So, before we perform higher order moments calculation in one wide centrality in heavy-ion collisions, the CBWC should be applied to address the effect of variation of the volume within a wide centrality bin. 
\begin{figure}[hptb]
\centering
\renewcommand{\figurename}{FIG.}
\vskip 0cm
\includegraphics[width=0.7\textwidth]{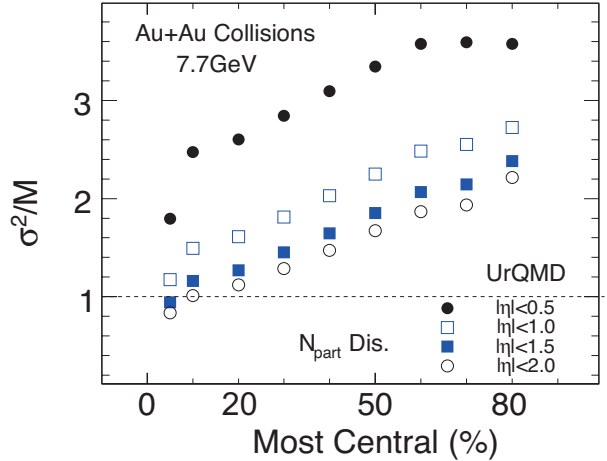}
\caption{(Color online) The centrality dependence of the $\sigma^2 / M$ of $N_{part}$ distributions for Au+Au collisions at \sNN=7.7 GeV in UrQMD model with different centrality definition.}
\label{Plot::R21}
\end{figure}

We also have tried to use the statistical errors as weight to perform the CBWC by replacing the weight factor $n_{r}$ in equ. (8)-(10) with $1/error^{2}$ for each single multiplicity bin. The statistical error is obtained by the Delta theorem method for moments of particle number distribution at each multiplicity bin. In Fig.~\ref{Plot::CBWC-sd} and~\ref{Plot::CBWC-kv} (open star), we find that the results with statistical error weighted CBWC are consistent with the results with events number weighted CBWC, but not for $\kappa\sigma^2$. It means the error weighted method can not be used for CBWC, which may be due to the statistical error is not only related to the number of events but also the moments itself.

\subsection{Centrality Resolution Effect (CRE)}
\label{Centrality reslution}
The definition of the centralities for two colliding nuclei is not unique and can be defined through different quantities. A frequently used quantity is the so called impact parameter $b$, defined as the distance between the geometrical centers of the colliding nuclei in the plane transverse to their direction. Other variables, such as the number of participant nucleons, $N_{part}$ and the number of binary collisions, $N_{coll}$, can be also used. Unfortunately, those geometrical variables can't be directly measured in the heavy-ion collision experiment and the collision centrality is determined from a comparison between experimental measures such as the particle multiplicity and Glauber Monte-Carlo simulations~\cite{Glauber}. Particle multiplicity, not only depends on the physics processes, but also reflects the initial geometry of the heavy-ion collision. This indicates that relation between measured particle multiplicity and impact parameter is not one-to-one correspond and there are fluctuations in the particle multiplicity even for a fixed impact parameter. Thus, it could have different initial collision geometry resolution (centrality resolution) for different centrality definitions with particle multiplicity. Fig.~\ref{Plot::R21} shows the centrality dependence of $\sigma^2/M$ of number of participant nucleons ($N_{part}$) distributions for Au+Au collisions at \sNN =7.7 GeV in UrQMD model. As the $N_{part}$ can reflect the initial geometry (volume) of the colliding nuclei, the $\sigma^2/M$ of $N_{part}$  distributions can be regarded as the centrality resolution for a certain centrality definition. It indicates that more particles are used in the centrality determination, the better centrality resolution and smaller fluctuation of the initial geometry (volume fluctuation) we get. This may affect moments of the event-by-event multiplicity distributions. 

\begin{figure}[thpb]
\centering
\renewcommand{\figurename}{FIG.}
\vskip 0cm
\includegraphics[width=0.7\textwidth]{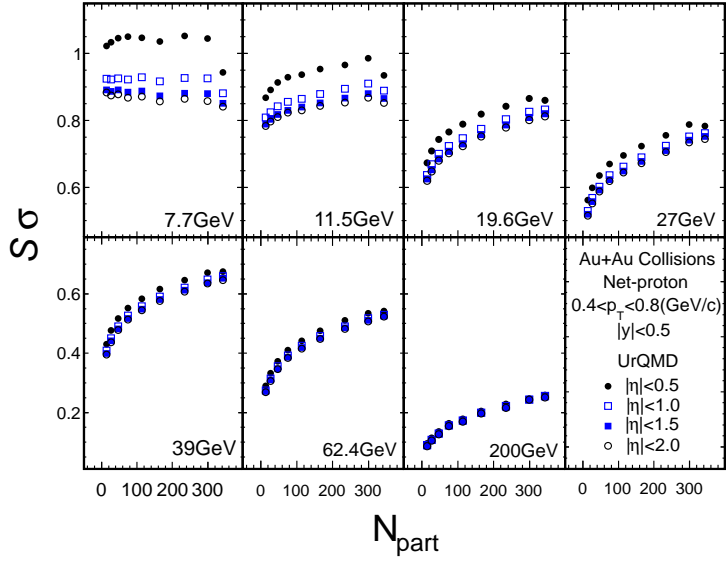}
\caption{(Color online) The centrality dependence of the moments
products $S\sigma$ of net-proton multiplicity distributions for Au+Au collisions at \sNN=7.7, 11.5, 19.6, 27, 39, 62.4, 200 GeV in UrQMD model
with different centrality definition.}
\label{Plot::res-sd}
\end{figure}

\begin{figure}[thpb]
\centering
\renewcommand{\figurename}{FIG.}
\vskip 0cm
\includegraphics[width=0.7\textwidth]{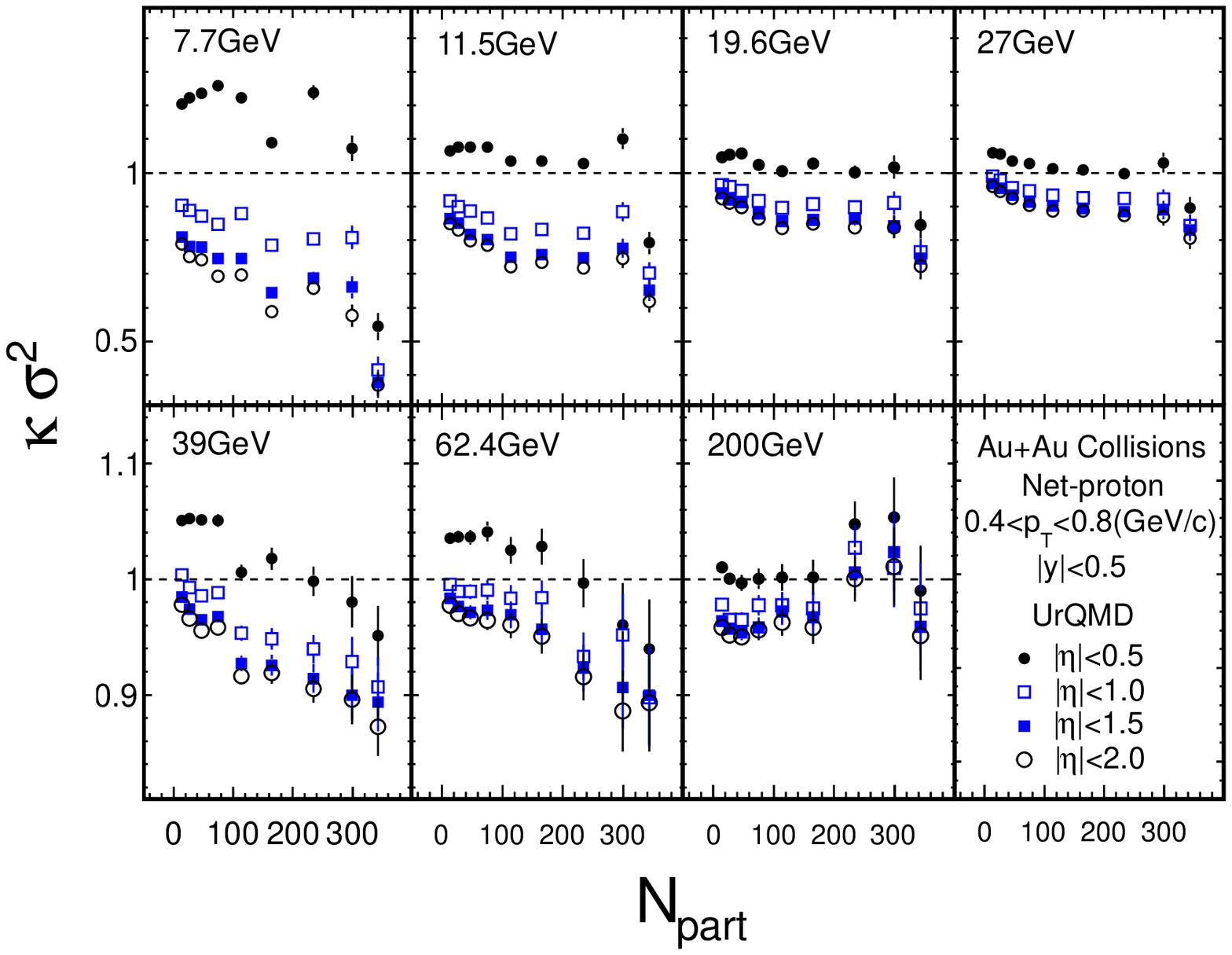}
\caption{(Color online) The centrality dependence of the moments
products $\kappa\sigma^2$ of net-proton multiplicity distributions for Au+Au collisions at \sNN=7.7, 11.5, 19.6, 27, 39, 62.4, 200 GeV in UrQMD model with different centrality definition.}
\label{Plot::res-kv}
\end{figure}

To verify CRE effect in the moment analysis, we use the charged kaon and pion multiplicity produced in the final state within $|\eta|<0.5, 1, 1.5$ and 2 to define the centrality in the UrQMD calculations. Fig.~\ref{Plot::res-sd} and~\ref{Plot::res-kv} show the centrality dependence of the moment products ($ S\sigma, \kappa\sigma^2$) of net-proton multiplicity distributions with four different $\eta$ ranges of charged kaon and pion used to determine the centrality. When we increase the $\eta$ range $(|\eta| < 1,1.5,2)$, the values of $S\sigma$ and $\kappa\sigma^2$ will decreases and show saturation around $|\eta|<2$, which indicates the centrality resolution effect will enhance the moments values of net-proton multiplicity distributions. The significant difference for moment products ($S\sigma, \kappa\sigma^2$) for the different $\eta$ range of the centrality definition can be understood as the different contribution from volume fluctuations arising from different centrality definition.

\begin{figure}[thpb]
\centering
\renewcommand{\figurename}{FIG.}
\vskip 0cm
\includegraphics[width=0.7\textwidth]{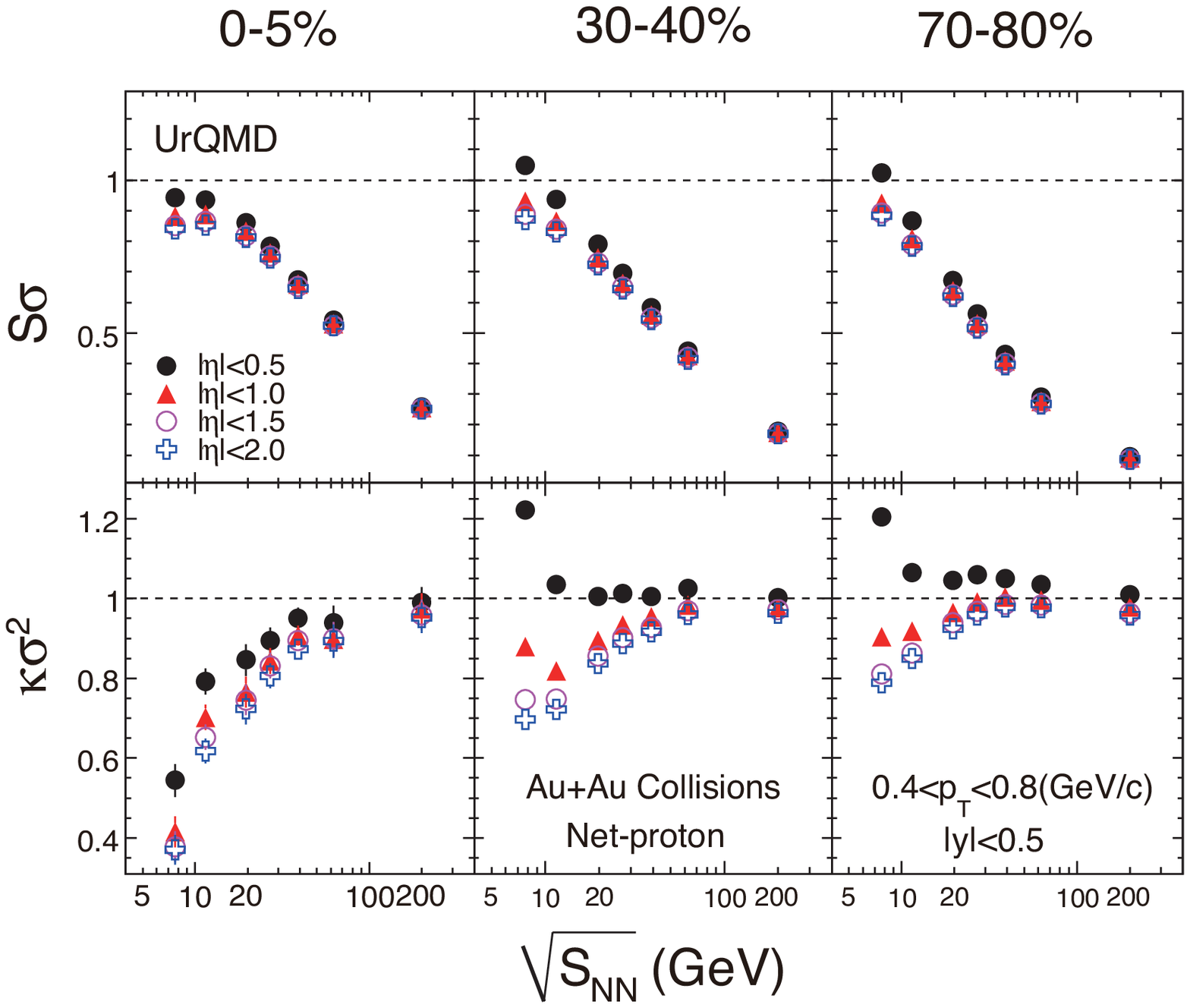}
\caption{(Color online) The energy dependence of the moments
products $(S\sigma , \kappa\sigma^2)$ of net-proton multiplicity distributions for Au+Au collisions at \sNN=7.7, 11.5, 19.6, 27, 39, 62.4, 200 GeV in UrQMD model with different centrality definition.}
\label{Plot::energy}
\end{figure}

Figure~\ref{Plot::energy} shows the energy dependence of moment product ($S\sigma, \kappa\sigma^2$) of net-proton multiplicity distributions for three different centralities (0-5\%, 30-40\%, 70-80\%) with four different $\eta$ ranges in Au+Au collisions. We can find that the $\kappa\sigma^2$ (fourth order fluctuation) is more sensitive to the centrality resolution effect than the {\SD} (third order fluctuation), and it has larger effect in the peripheral collision as well as at low energies. Thus, we should use a lager $\eta$ range in the centrality definition for the real experimental moment analysis. 

\begin{figure}[thpb]
\centering
\renewcommand{\figurename}{FIG.}
\vskip 0cm
\includegraphics[width=0.7\textwidth]{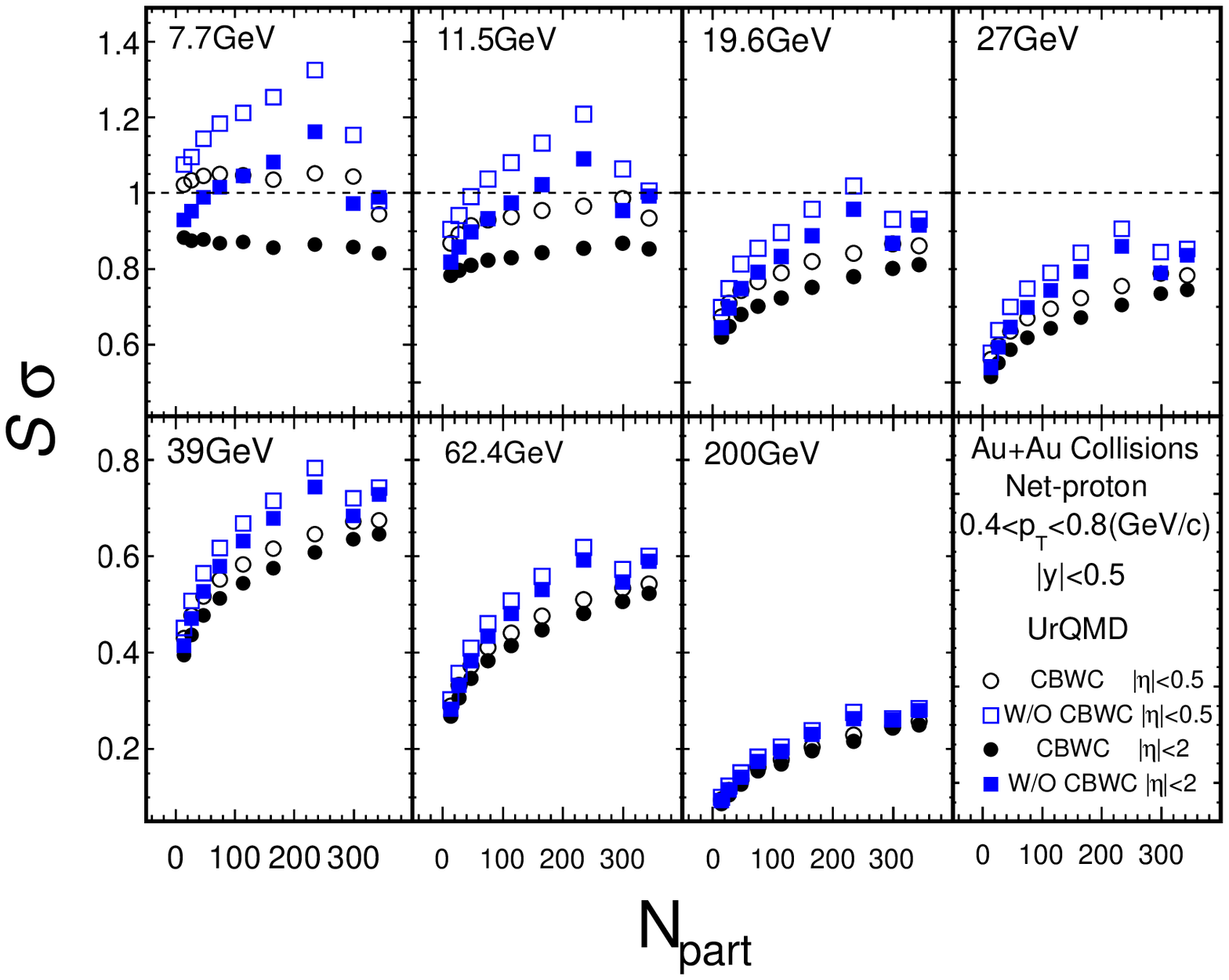}
\caption{(Color online) The centrality dependence of the moments
products $S\sigma$ of net-proton multiplicity distributions for Au+Au collisions at \sNN=7.7, 11.5, 19.6, 27, 39, 62.4, 200 GeV in UrQMD model. The open circle and square are the results with CBWC and without CBWC at $|\eta|<0.5$ for centrality definition, respectively. The solid circle and square are the results with CBWC and without CBWC at $|\eta|< 2$ for centrality definition, respectively.}
\label{Plot::res-cbwc-sd}
\end{figure}

\begin{figure}[thpb]
\centering
\renewcommand{\figurename}{FIG.}
\vskip 0cm
\includegraphics[width=0.7\textwidth]{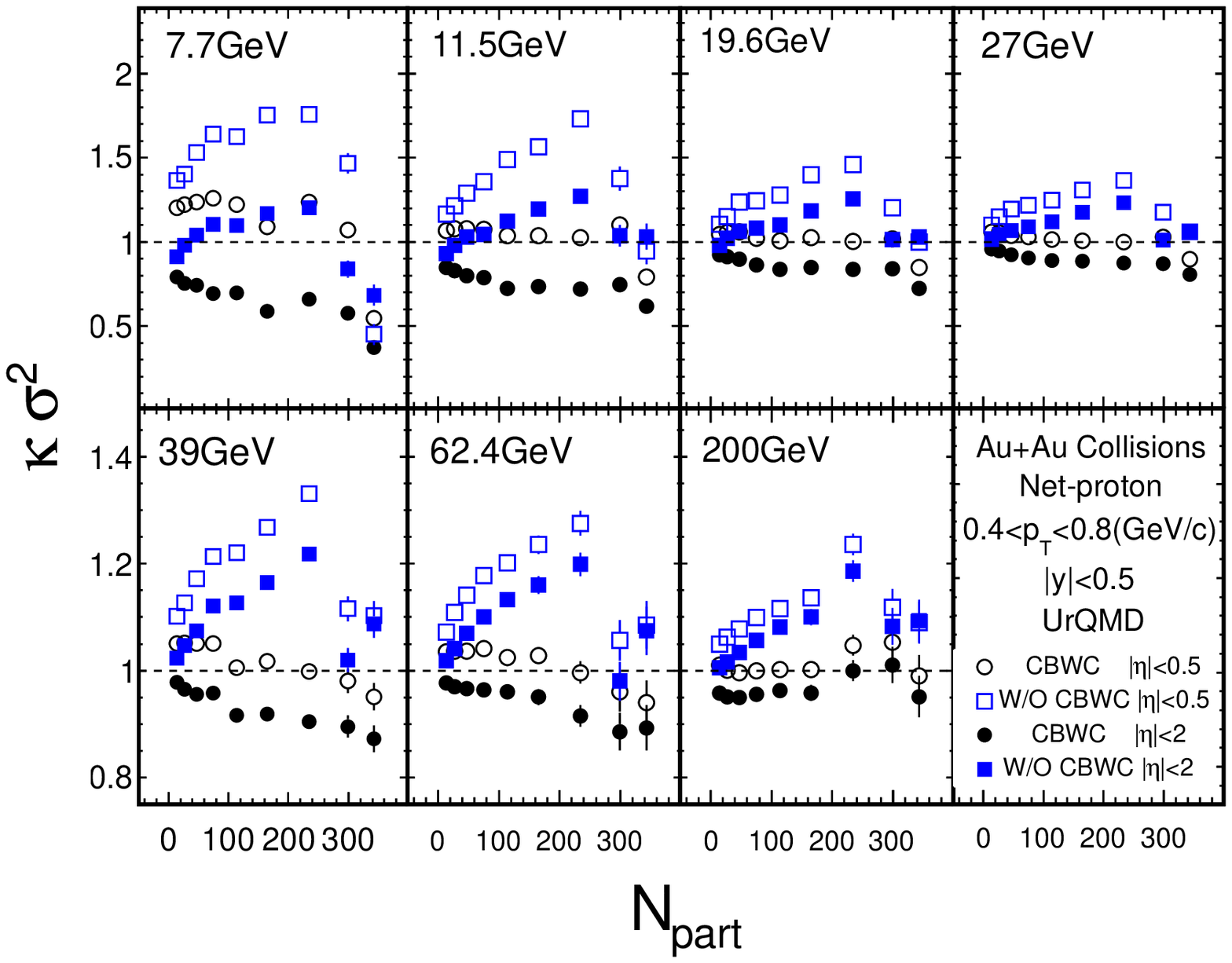}
\caption{(Color online) The centrality dependence of the moment
products $\kappa\sigma^2$ of net-proton multiplicity distributions for Au+Au collisions at \sNN=7.7, 11.5, 19.6, 27, 39, 62.4, 200 GeV in UrQMD model.
The open circle and square are the results with CBWC and without CBWC at $|\eta|<0.5$ for centrality definition. The solid circle and square are the results with CBWC and without CBWC at $|\eta|< 2$ for centrality definition.}
\label{Plot::res-cbwc-kv}
\end{figure}

Figure~\ref{Plot::res-cbwc-sd} and~\ref{Plot::res-cbwc-kv} show the centrality dependence of moment products $(S\sigma, \kappa\sigma^2)$ of net-proton multiplicity distributions for Au+Au collisions with CBWC and without CBWC at different centrality resolution $(|\eta| < 0.5,2)$.  We can see that 
the CBWE and CRE have larger effect in the moment products at low energies than at high energies. The magnitude of CBWE in the moment products is comparable with the magnitude of CRE at low energies, while the CBWE has larger effects at higher energies. 

Let's summarize the CBWE and the CRE in the moment analysis in heavy-ion collisions. The CBWE stems from the variation of the volume when we perform the moment calculations within one wide centrality bin and the CRE is originated from the initial volume (geometry) fluctuations of the colliding nuclei. Both of those two effects are resulted from the volume fluctuation effects and it can be effectively addressed by using the CBWC and more particles (larger $\eta$ coverage) in centrality definition, respectively.
 
\begin{figure}[thpb]
\centering
\renewcommand{\figurename}{FIG.}
\vskip 0cm
\includegraphics[width=0.7\textwidth]{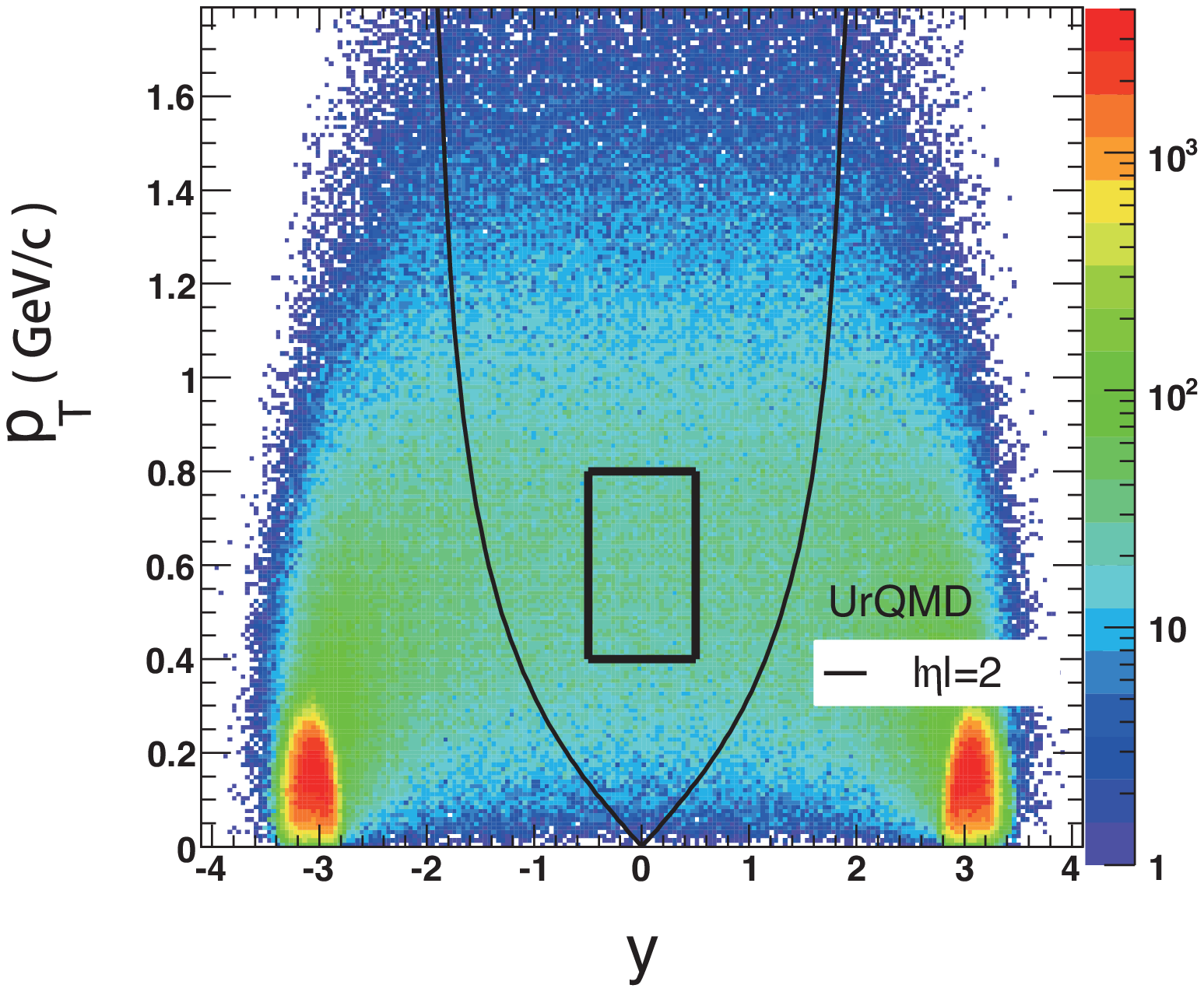}
\caption{(Color online) The proton distribution in the phase space of transverse momentum ($p_T$ GeV/c) and rapidity $(y)$ for Au+Au collisions at \sNN =19.6 GeV using the UrQMD model. The relation between $p_{T}$ and $y$ is ${p_T} = m_{0}/\sqrt {{{\sinh }^2}\eta /{{\sinh }^2}y - 1}$, where the $m_{0}$ is the particle rest mass.}
\label{Plot::y-pt}
\end{figure}

\begin{figure}[thpb]
\centering
\renewcommand{\figurename}{FIG.}
\vskip 0cm
\includegraphics[width=0.7\textwidth]{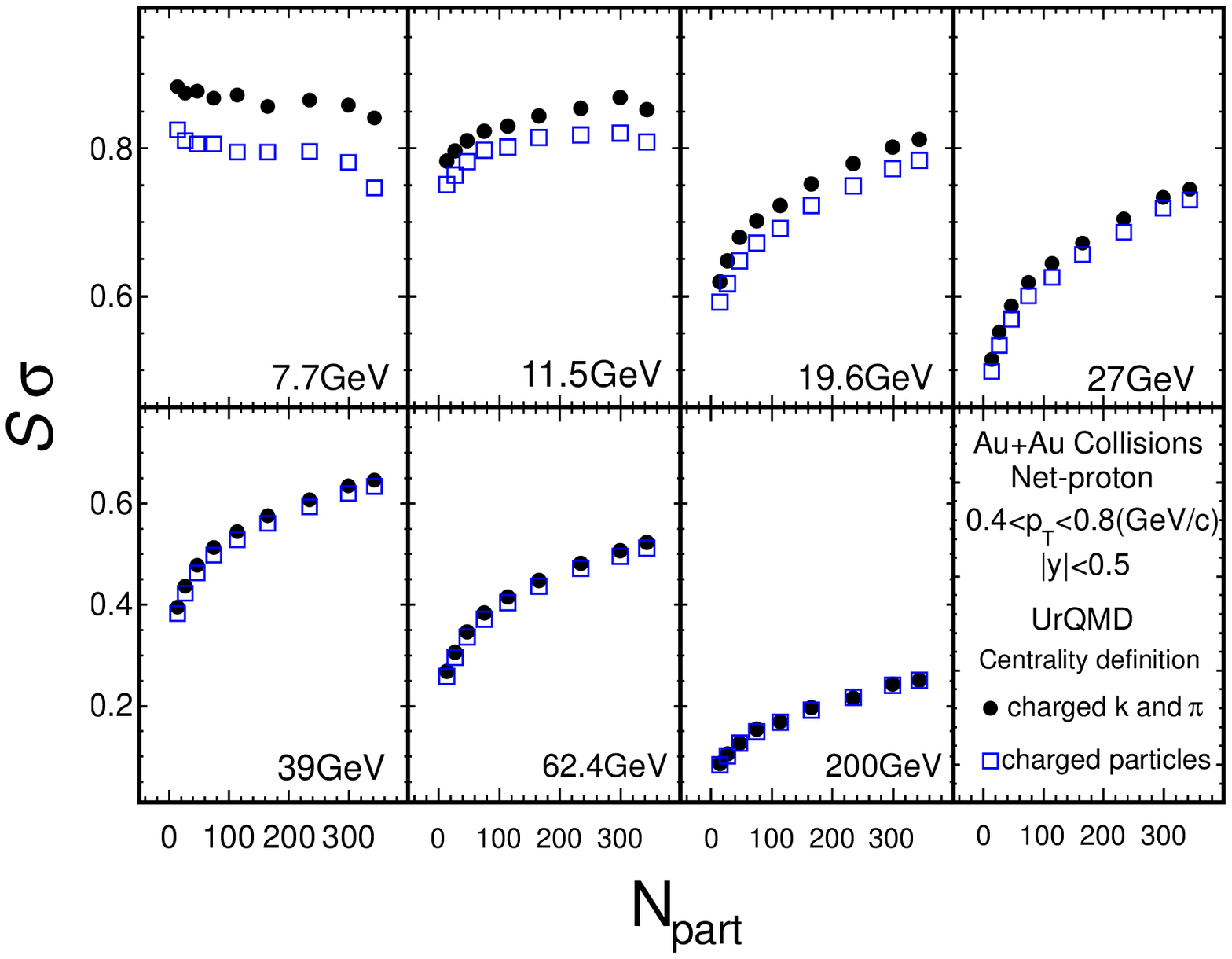}
\caption{(Color online) The centrality dependence of the moments
products $S\sigma$ of net-proton multiplicity distributions for Au+Au collisions at \sNN=7.7, 11.5, 19.6, 27, 39, 62.4, 200 GeV in UrQMD model with
two type of centrality definition.}
\label{Plot::ac-sd}
\end{figure}

\begin{figure}[hptb]
\centering
\renewcommand{\figurename}{FIG.}
\vskip 0cm
\includegraphics[width=0.7\textwidth]{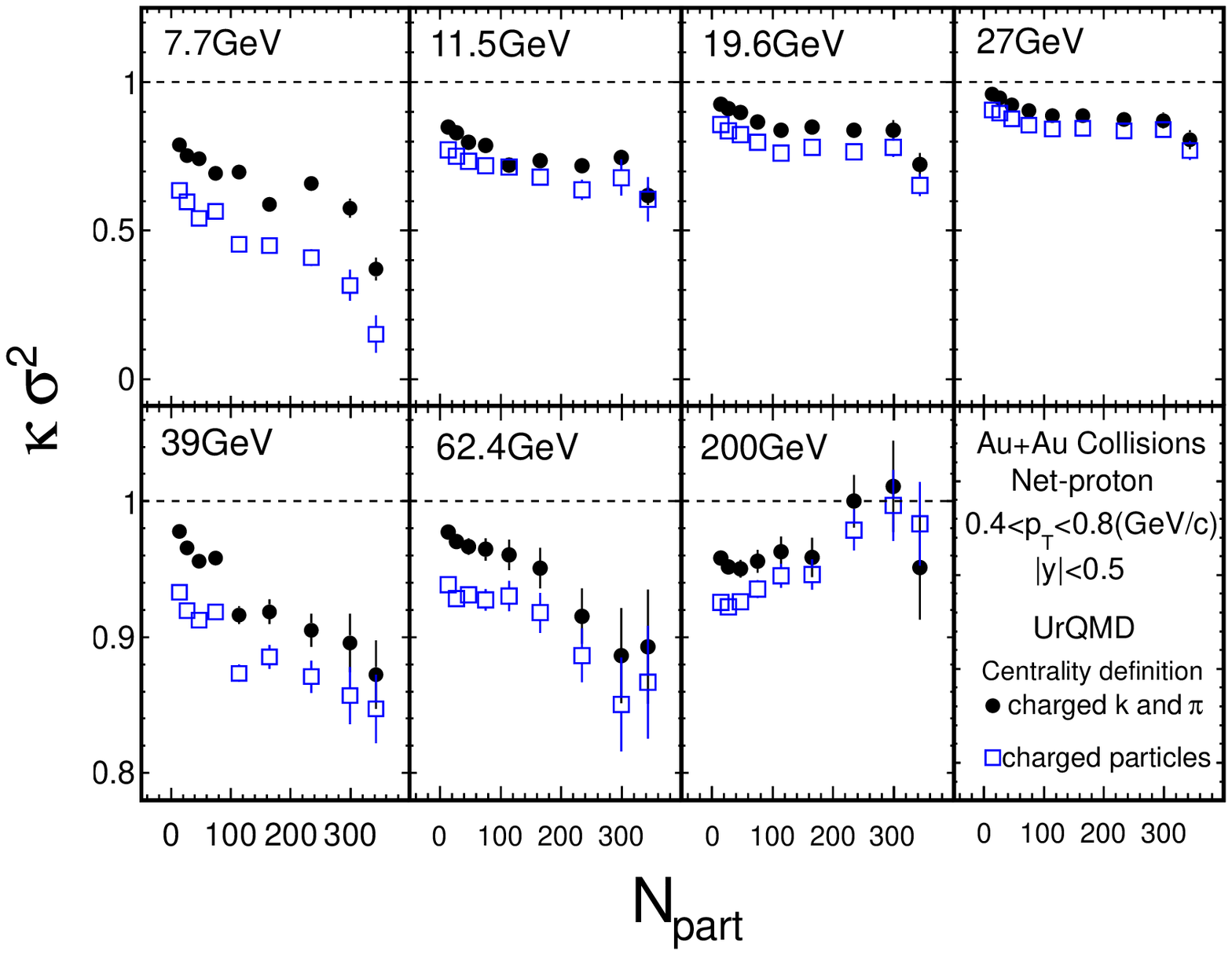}
\caption{(Color online) The centrality dependence of the moments
products $\kappa\sigma^2$ of net-proton multiplicity distributions for Au+Au collisions at \sNN=7.7, 11.5, 19.6, 27, 39, 62.4, 200 GeV in UrQMD model
with two type of centrality definition.}
\label{Plot::ac-kv}
\end{figure}

\section{Auto-correlation Effect (ACE)}
\label{auto-correlation}
In the previous section, we use the multiplicity of charged kaon and pion to define the centrality in Au+Au collisions. 
The reason is to avoid the effect of auto-correlation between protons/anti-protons involved in our moments analysis and in the centrality definition.  Fig.~\ref{Plot::y-pt} shows the proton distribution in the phase space of transverse momentum $(p_T)$ and rapidity $(y)$ for Au+Au collisions at \sNN=19.6 GeV in the UrQMD model. The protons used in the moment analysis are distributed in the oblong box ($0.4<p_{T}<0.8 $ GeV/c, $|y| < 0.5$). The solid black curve represents ($p_{T}$, $y$) points with $\eta=2$. If all the charged particles within the $\eta=2$ curve are used to define the centrality, it will definitely introduce the auto-correlation effects.

Figure~\ref{Plot::ac-sd} and~\ref{Plot::ac-kv} show the centrality dependence of the moment products $S\sigma$ and $\kappa\sigma^2$ of net-proton distributions with different centrality definition, respectively. The difference between the results from the two types of centrality definition is larger in low energies than high energies. This is because the number of protons commonly used in the moments analysis and the centrality definition
are larger in low energies than high energies. The auto-correlation effect will lead to the suppressing of the value of the moment products $S\sigma$ and $\kappa\sigma^2$. Thus, to avoid the auto-correlation effect, we should exclude the corresponding proton/anti-protons from the centrality definition.

\section{Summary}
\label{sect_summary}
We presented three measurement artifacts in the moment analysis of net-proton distributions in heavy-ion collisions. Those issues could be commonly encountered in any moment analysis of multiplicity distributions in heavy-ion collisions. Those are centrality bin width effect, centrality resolution effect and auto-correlation effect. We also discussed corresponding methods to address those effects to obtain physics information. The studies were done using events from UrQMD model and the calculations are carried out for Au+Au collisions at \sNN = 7.7, 11.5, 19.6, 27, 39, 62.4, 200 GeV. The centrality bin width effect and centrality resolution effect are originated from the system volume and its fluctuations. The auto-correlation effect is resulted from using the same particles for the moment analysis and for the centrality definitions. We have also discussed the statistical error estimation method used in moments analysis. The effects and techniques discussed in this paper are not only restricted within the net-proton moment analysis, but also applicable to other moment analysis in heavy-ion collisions, such as net-charge, net-strangeness. This will provide a reference on how to perform model calculation with similar analysis techniques for comparison with the experimental results.

\section*{Acknowledgement}
The work was supported in part by the National Natural Science
Foundation of China under grant No. 11205067 and 11135011. CCNU-QLPL Innovation Fund(QLPL2011P01, QLPL2013P01)
and China Postdoctoral Science Foundation (2012M511237).  BM is supported by the DST
SwarnaJayanti project fellowship.

\section*{References}
\bibliography{techniques}
\bibliographystyle{unsrt}
\end{document}